# Online and Offline Evaluation in Search Clarification

LEILA TAVAKOLI, RMIT University, Australia

JOHANNE R. TRIPPAS, RMIT University, Australia

HAMED ZAMANI, University of Massachusetts Amherst, United States

FALK SCHOLER, RMIT University, Australia

MARK SANDERSON, RMIT University, Australia

The effectiveness of clarification question models in engaging users within search systems is currently constrained, casting doubt on their overall usefulness. To improve the performance of these models, it is crucial to employ assessment approaches that encompass both real-time feedback from users (online evaluation) and the characteristics of clarification questions evaluated through human assessment (offline evaluation). However, the relationship between online and offline evaluations has been debated in information retrieval. This study aims to investigate how this discordance holds in search clarification. We use user engagement as ground truth and employ several offline labels to investigate to what extent the offline ranked lists of clarification resemble the ideal ranked lists based on online user engagement. Contrary to the current understanding that offline evaluations fall short of supporting online evaluations, we indicate that when identifying the most engaging clarification questions from the user's perspective, online and offline evaluations correspond with each other. We show that the query length does not influence the relationship between online and offline evaluations, and reducing uncertainty in online evaluation strengthens this relationship. We illustrate that an engaging clarification needs to excel from multiple perspectives, and SERP quality and characteristics of the clarification are equally important. We also investigate if human labels can enhance the performance of Large Language Models (LLMs) and Learning-to-Rank (LTR) models in identifying the most engaging clarification questions from the user's perspective by incorporating offline evaluations as input features. Our results indicate that Learning-to-Rank models do not perform better than individual offline labels. However, GPT, an LLM, emerges as the standout performer, surpassing all Learning-to-Rank models and offline labels.

## 1 INTRODUCTION

When a user submits a query to a search engine like *Bing*, in addition to the results page, the search engine sometimes presents a multi-choice clarification question. This clarification question aims to help users specify their information needs. Although multiple clarification questions can be generated for a single query, only one is typically presented to the user. Despite the advancements in generating clarification questions in search systems, the success rate of users engaging with such clarification questions remains low [74]. An analysis of the largest search clarification dataset,

---

Authors' addresses: Leila Tavakoli, RMIT University, Australia, leila.tavakoli@rmit.edu.au; Johanne R. Trippas, RMIT University, Australia, j.trippas@rmit.edu.au; Hamed Zamani, University of Massachusetts Amherst, United States, zamani@cs.umass.edu; Falk Scholer, RMIT University, GPO Box 2476, Australia, falk.scholer@rmit.edu.au; Mark Sanderson, RMIT University, Australia, mark.sanderson@rmit.edu.au.

*MIMICS* [74], demonstrates that users tend to engage more with certain clarification questions than others for a given query. Furthermore, many clarifications are left unengaged, regardless of how many times they are presented to users (e.g., only about 17% of query-clarification pairs in the *MIMICS-Click* dataset, a subset of the *MIMICS* dataset, received positive engagement). This indicates that users are not easily engaged with clarification questions, and clarifications are not equally engaging from users' perspectives raising questions about the overall effectiveness of the search clarification question models.

An engaging clarification question should encourage users to actively participate in the search process and interact with the system. This interaction can lead to a more personalised and satisfying search experience and save time by quickly guiding them toward relevant results [72, 78]. User engagement has emerged as a crucial metric in interactive information retrieval studies. This is particularly significant for both commercial entities like search engines and e-commerce businesses, as well as educational institutions such as libraries, who are now placing emphasis on acquiring and keeping their customers [47]. To attain a high level of user engagement for a clarification model, it is essential to employ evaluation techniques that consider both user behaviour and the characteristics of engaging clarification questions. The typical evaluation process in deploying new models in search engines involves *(1) offline evaluation* with labelled test collections and *(2) online evaluation* through user interactions, often using A/B testing. A reliable offline evaluation dataset is crucial for continuous research iterations and the refinement of models and features. Researchers commonly base their online experiments on findings from offline evaluations due to the resource-intensive nature of online assessments. However, the relationship between offline and online evaluations in search clarifications is relatively unexplored. For example, Zamani et al. [73] introduced three distinct models for generating clarification questions in an open-domain information-seeking system. Nevertheless, the evaluation of these models' performance relied solely on human annotation, without investigating how they perform in real-world conditions. To bridge this knowledge gap, we investigate the relationship between user engagement (online evaluation) and the characteristics of clarifications that are manually evaluated (offline evaluation) by studying the following two primary research questions:

- **RQ1**: How well do offline evaluations correspond with online evaluations in search clarification?

Following the study conducted by Zamani et al. [74], we focus on clarification panes, each consisting of a clarification question and up to five candidate answers. Figure 1 shows an example of a clarification pane presented to users on the *Bing* search engine. The ground truth in this study is the ideal ranked list of clarification panes generated based on user engagement. An ideal ranked list of clarification panes is a list that has the most engaging clarification pane (MECP) at the first position, and the rest of the clarification panes are sorted based on the *Engagement Level* in descending order. We aim to determine two aspects: *(i)* whether the offline labels can successfully position MECP at the top of the ranked list, and *(ii)* to what extent the ranked lists generated by the offline labels resemble the ideal ranked lists for clarification panes. We initially evaluate the effectiveness of an oracle[1] clarification selection model. This model has access to every offline label, and its performance in terms of the similarity of generated ranked lists with the ideal ranked lists is evaluated. Offline labels are different characteristics of clarification panes such as quality, coverage, diversity and importance order of candidate answers annotated by the human judgement, and the online label is real user engagement level. The details of the labels will be discussed in Section 3. We move beyond the assumption that the offline labels are independent of each other and delve into their combination by utilising Learning-to-Rank (LTR) models to determine if these combinations align better with online evaluation. Additionally, we use a large language model (LLM) to predict online user engagement with clarification, considering the provided offline labels as the input for the model. Motivated by Zamani et al. [75], who showed user behaviour is different in short queries (often keyword queries) and long queries (often natural language questions), we further investigate the impact of query length on the relationship between online and offline evaluations in search clarification.

---

[1]In machine learning, an oracle typically refers to an idealised entity or concept that provides perfect information or answers to a given problem. It is often used as a theoretical reference point to establish performance bounds or to measure the efficiency and effectiveness of an algorithm.

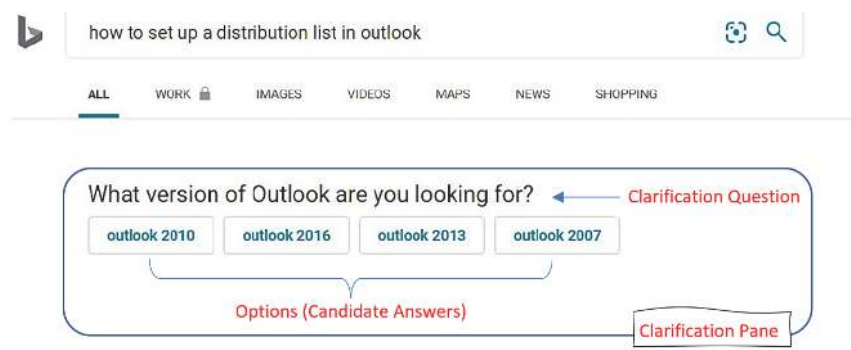

Fig. 1. A clarification pane shown after a user query [74].

Uncertainty in collected online evaluations, much like in any form of assessment, has far-reaching implications. It not only undermines the trustworthiness of the online evaluation results and the inferences that can be drawn from them, but it also introduces a potential variable that can disrupt the alignment between online and offline evaluations. This inquiry is pivotal to shed light on strategies to mitigate the impact of uncertainty in online assessments. To examine this phenomenon, we aim to address the following research question:

- **RQ2**: How does uncertainty in the online evaluation impact the relationship between online and offline evaluation?

Here, we control uncertainty in the online labelling based on the number of times a clarification question is presented to users, known as *Impression Level*. The higher the *Impression Level*, the more reliable (thus less uncertain) online labels based on click-through rate are.

In contrast to the widely held notion that online and offline evaluations do not always coincide regarding retrieval quality [16, 18, 22, 22, 56], our study shows that offline evaluations align with online evaluations in search clarification. However, certain essential factors should be considered. This study also enhances our comprehension of the performance of LLMs in predicting online user engagement with clarifications when offline labels are employed as input for the models. The insights gained from our investigation will aid in refining the evaluation methodology for search clarification, resulting in improved user search experiences and more effective decision-making when implementing clarification models.

## 2   RELATED WORK

We present a summary of previous works on clarification questions and online and offline evaluations in information retrieval.

## 2.1 Search Clarification

The use of clarification questions to improve user satisfaction has been investigated in different areas such as search engines [75], conversational search systems [39], chat bots [52], question-answering forums [62], and spoken dialogue systems [21]. Generating and selecting clarification questions, two areas of interest [10, 39, 62], are discussed here, and they are followed by a summary of available search clarification datasets.

*2.1.1* ***Clarification question generation***. Clarification question generation is a relatively new research area in information retrieval. In 2019, Rao and Daumé III [54] proposed an adversarial training approach for generating clarification questions. Their study inspired further research by Zamani et al. [73] and Shwartz et al. [58], who focused on designing clarification systems. Zamani et al. [73] explored generating clarification questions for open-domain search by proposing three different models. Shwartz et al. [58] proposed an unsupervised framework using self-talk to generate natural language clarification questions and answers. The evaluation of these models primarily relied on offline human judgements, leaving a knowledge gap regarding their performance from an online user's perspective.

*2.1.2* ***Clarification question selection***. Several studies investigated the clarification question selection. Rao and Daumé III [53] developed a neural network model that taught machines to ask clarification questions in uncertain situations. Aliannejadi et al. [5] explored asking clarification questions in open-domain information-seeking conversational systems. They showed that their model outperformed baselines and improved user satisfaction. Ou and Lin [48] proposed a clarification question selection system for recalling and ranking such questions. Kumar et al. [40] investigated asking clarification questions in *StackExchange* and demonstrated the high performance of BERT representations on this task. Recent works by Sekulić et al. [57] and Zamani et al. [75] have further contributed to the development of clarification question selection systems, focusing on response understanding, user interaction analysis, and user engagement prediction.

*2.1.3* ***Search clarification datasets***. Several search clarification datasets have been created over the last few years [3–5, 51, 69, 74]. For example, Xu et al. [69] created CLAQUA, a clarification dataset of 40,000 open-domain examples to enable systems to ask clarification questions in open-domain question answering. This dataset supported three tasks: given a question, check whether clarification is needed; if yes, generate a clarification question; and then predict answers based on user feedback. Aliannejadi et al. [5] collected a clarification dataset through crowd-sourcing named Qulac. This dataset was built on top of the TREC Web Track 2009-2012 data and contained over 10,000 question-answer pairs for 198 TREC topics with 762 facets. Inspired by Qulac, Aliannejadi et al. [3, 4] crowd-sourced new datasets to study clarification questions that were suitable for conversational settings and in open domain dialogues focusing on single and multi-turn conversations. Penha et al. [51] created a dataset that focused on the interaction between an agent and a user, including clarification questions. The researchers presented a conceptual model and provided baseline results for conversation response ranking and user intent prediction tasks.

The largest search clarification dataset, *MIMICS*, was introduced by Zamani et al. [74] and was extracted from *Bing* search engine. Each clarification was generated by a *Bing* production algorithm and contained a clarification question and up to five candidate answers. Compared to other datasets, *MIMICS* contains realistic queries and user interaction signals and covers many clarification types. *MIMICS* also contains search engine results pages (SERPs) of up to ten retrieved documents, including a title, URL, and snippet for each query. The *MIMICS* data collection consists of three datasets of *MIMICS-Click*, *MIMICS-ClickExplore*, and *MIMICS-Manual*.

The most recent search clarification dataset, built as an extension of the pre-existing *MIMICS-ClickExplore* dataset, was called *MIMICS-Duo* and introduced by Tavakoli et al. [61]. It contains 306 unique queries with multiple clarification panes (1,034 query-clarification pairs), interactions of real users, and graded quality labels including multiple clarification panes rating, overall quality labelling for clarification panes and their candidate answers and labels for different aspects of clarification panes. Contrary to other search clarification datasets, *MIMICS-Duo* contains online and offline evaluations created through crowd-sourcing. This dataset enables us to analyse the relationship between the online and offline evaluations in search clarification, addressed in the current publication.

### 2.2 Online and Offline Evaluation Approaches

To understand what makes a clarification question engaging from a user's point of view, the relationships between various characteristics of the clarification questions, labelled by human judgement, and explicit user interaction, known as user engagement, need to be investigated. Such studies are known as online–offline evaluations, and we review the previous works on this topic now.

There are two approaches in general to evaluate retrieval quality: *(i)* manual judgements of the relevance of documents to queries provided by trained annotators (offline evaluation) [15] and *(ii)* user behaviour observations when presenting the search results (online evaluation) [11]. While offline evaluations are performed on pre-collected datasets, online evaluations involve testing the system in real-time using actual users. Both approaches have advantages and disadvantages, and the choice of which method to use depends on various factors, such as the type of system being evaluated and the available resources. The effectiveness of using human judgements in quality retrieval analysis has been demonstrated before [65]. Offline evaluations are often used before deploying new ranking policies, which help to run A/B testing[2] more safely and intelligently [14, 41]. However, such an evaluation has two limitations. First, human annotations may not be capable of reflecting the actual relevance and cannot reliably estimate the user's actual information need simply based on the query issued and inaccurately reflect user utility [1, 12]. This comes from the fact that different users may issue the same textual query with different information needs or intents [63]. Moreover, It was understood that users' emotion control (EC) interacts with search tasks and influences the search behaviour which may not be captured by the annotators [36]. Second, the cost of conducting offline evaluations, such as hiring annotators or setting up infrastructure, is typically substantial. Additionally, offline evaluations usually take considerable time to complete, ranging from days to weeks or even longer. These factors limit offline evaluations' benefits for many organisations or projects, as the expenses and time required may be too burdensome. Consequently, alternative, more cost-effective, faster evaluation methods, such as online evaluations, are often preferred. These online metrics are based on observable user behaviour [11, 34] and include: Click Through Rate (CTR) and the ranks of clicked documents [31] as well as their extensions (e.g., A binary value representing click) [14], Precision at Lowest Click (PLC) (i.e., number of clicks divided by the position of the lowest click) [23]), dwell time including query dwell time, time to first click, the average of click dwell time [28, 71], query reformulations, response times, how the session was terminated (e.g., by closing the browser window or by typing a new Internet address) [20], mouse movement and per-topic reading time [37]. Online evaluations can be grouped into two classes of absolute metrics and pairwise preferences [45]. Contrary to absolute metrics that provide an overall assessment of the retrieval performance based on predefined criteria, pairwise preference methods such as interleaving assume that the better of two (or more) options can be identified based on user behaviour. For example, clicked results are preferred over results previously skipped in the ranking [33]. Despite the enormous value of click-through data, it is inherently biased and very noisy [66]. There are multiple sources of bias, including position bias [32], presentation bias (e.g., the position of results in the ranking) [60], and trust bias [49]. Such noisy data may lead to biased training data that negatively affects the downstream applications [29]. There are also some other factors, such as educational level, intelligence, and familiarity with Information Retrieval systems that impact the decision of user satisfaction and the click-through data [2, 26, 42] making the data difficult to interpret. This agrees with observations by Zheng et al. [77] that click-through data and relevance do not always correlate and CTR should be used with precaution.

---

[2]A randomised experiment that usually involves two variants (A and B), shown to users, and statistical analysis is used to determine which variation performs better) [38].

Substantial discrepancies between the offline and online evaluations have been reported in the literature. Cremonesi et al. [16], Ekstrand et al. [18], Garcin et al. [22], Said and Bellogín [56] identified several inconsistencies when investigating recommendation methods using online and offline evaluations. Yi et al. [70] investigated the performance of predictive models for search advertising using online and offline evaluation metrics and showed that some offline metrics like AUC (the Area Under the Receiver Operating Characteristic Curve) and RIG (Relative Information Gain) could be misleading and result in a discrepancy in online and offline metrics. Such discrepancy was also observed and stated by Beel et al. [8] and Beel and Langer [7]. In another study, Garcin et al. [22] investigated news recommenders and showed that in an offline setting, recommending popular stories is a winning strategy, but in an online setting, it was the poorest.

Online evaluations can also be misleading. Zheng et al. [77] and later Garcin et al. [22] showed that CTR, an adopted and widely accepted metric in online evaluations, overestimates the impact of popular items. In fact, recommending items with higher CTR does not necessarily imply higher relevance of two items, and factors like item popularity, item serendipity or the placement/order of recommendations may also influence a user's click behaviour.

Chen et al. [13] conducted a meta-evaluation of a series of existing online and offline metrics to study how well they predict actual search user satisfaction in different search scenarios. They showed both types of evaluation noticeably correlate with user satisfaction, but they reflect satisfaction from different perspectives and for different search tasks. They observed a strong correlation between top-weighted offline metrics and user satisfaction in homogeneous search (i.e. ten blue links), whereas online metrics outperform offline metrics when vertical results are federated. They also understood that incorporating mouse hover information into existing online evaluation metrics better aligns with search user satisfaction than click-based online metrics. Liu and Yu [43] believed users often seek different goals at different search moments, which may evaluate system performances differently. Therefore, achieving real-time adaptive search evaluation and recommendation would be difficult. They meta-evaluated a series of online and offline evaluation metrics through a user study. Their results showed that the performance of query-related and online features had large variations across different task states. However, offline evaluation metrics generally had stronger correlations with user satisfaction. In another study, Rossetti et al. [55] showed that with the same set of users, the ranking of algorithms based on offline accuracy measurements contradicts the results from the online study. Later, a comparison of online and offline assessments for Query Auto Completion was carried out by Bampoulidis et al. [6], and it showed a large potential for significant bias if the raw data used in an online experiment is re-used for offline evaluations. It is worth noting that a lack of correlation between offline and online evaluations in voice shopping traffic and Web image search was also reported by Zhang et al. [76] and Ingber et al. [27].

While prior works have offered insight into how well online and offline evaluations correlate in retrieval quality, there is no extensive study on this controversial topic in search clarification. The only available study was conducted by Zamani et al. [74], who examined the *MIMICS* dataset and investigated correlations between online and offline evaluations using a single offline label. They concluded that no correlation was observed between the two evaluation methods. The focus of our study is to investigate the relationship between online and offline evaluations in terms of ranking multiple clarification panes and identifying the most engaging clarification pane for a given query. Next, we group the query-clarification pairs based on the query length and *Impression Level* for a more detailed study. Furthermore, we investigate if the combination of offline labels aligns better with the online label using a series of LTR models. Finally, the performance of an LLM in predicting user engagement with and without incorporating the offline labels as the model input is studied.

## 3 METHODOLOGY

First, we describe the dataset used in our experiments in Section 3.1, including the online and offline labels. We then explain the experimental design in Section 3.2, including our approach to investigating the relationship between the online and offline evaluations. Finally, we specify the evaluation metrics used in Section 3.3.

### 3.1 Dataset

In this study, we use the *MIMICS-Duo* dataset that contains both online and offline evaluations for 1,034 query-clarification pairs. To ensure the accuracy of the collected labels, Tavakoli et al. [61] conducted extensive quality assurance and attention measures in addition to pilot surveys, which led to a success rate of higher than 90% for the data collection. The dataset details and labels used in this study are now discussed.

*3.1.1* **Online labels**. Online labels in the *MIMICS-Duo* dataset include *Engagement Level* and *Impression Level*. The *Engagement Level* is constructed based on the click-through rate of real user interactions with clarification panes in Bing [74]. An equal-depth method was used for *Engagement Level*, dividing all the positive click-through rates into ten bins. Hence, the *Engagement Level* is an integer between 1 to 10 presenting the level of total engagement received by users in terms of click-through rate. Moreover, an *Engagement Level* of 0 was assigned to clarification panes with no clicks. According to Tavakoli et al. [61], collected queries have different topics and intents, and they attempted to keep a balance between the number of query-clarification pairs with different *Engagement Levels*. The second online label is the *Impression Level*, computed based on the number of times a given query-clarification pair was presented to users. Every query-clarification pair in the dataset was shown at least ten times to search engine users. The *Impression Level* has three quality values (low, medium, and high) and correlates with the query frequency. This study uses this online label to group the clarification panes for the experiments in Subsection 4.2.

*3.1.2* **Offline labels**. Offline labels in the *MIMICS-Duo* dataset include a series of crowd-sourcing labels consisting of *(i) List-wise Preference*, *(ii) Quality Labelling*, and *(iii) Aspect Labelling*.

The *List-wise Preference* was collected based on crowd-sourced worker preferences. Workers were simultaneously shown all generated clarification panes (varied between three to eight depending on the query) for a given query. They were asked to rate the clarification panes using a 5-point rating (five means highest preference, and one means lowest preference). The nature of this label is different from other labels. For this label, all clarification panes for a given query were relatively rated with respect to each other at the same time. However, for the other two labelling tasks, workers were shown one clarification pane and asked to annotate only one characteristic of the clarification pane in isolation.

The *Quality Labelling* consists of two quality measures, the *Overall Quality* of the complete clarification panes and *Option Quality*, that is, the quality of individual options (clarification pane candidate answers). Crowd-source workers rated the clarification panes and the quality of their options with a 5-point rating (five means very good quality, and one means very bad quality).

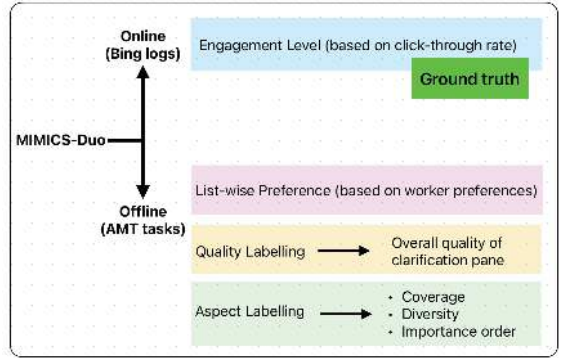

Fig. 2. An overview of variables used in this study from the *MIMICS-Duo* dataset.

*Aspect Labelling* consists of four sub-labels, that is, *Coverage* (i.e., the extent to which the clarification pane covers every potential aspect of the query), *Diversity* (i.e., the extent to which the clarification pane does not contain redundant information), *Understandability* (i.e., the extent to which the clarification pane is digestible and meaningful), and *Importance Order* (i.e., the extent to which the most relevant and important candidate answers are positioned from left to right). Workers were asked to label a clarification pane for these aspects through a 5-point rating (e.g., five means the worker strongly agreed that the clarification pane had high coverage, and one means the worker strongly disagreed that the clarification pane had a high coverage).

### 3.2 Experimental Design

We showed that each clarification pane has two types of labels, online and offline. We use one online label (i.e., *Engagement Level*) and five offline labels (i.e., *List-wise Preference*, *Overall Quality*, *Coverage*, *Diversity*, and *Importance Order*) to investigate the relationship between online and offline evaluations in search clarification. In the *MIMICS-Duo* dataset, *Overall Quality* and *Option Quality* labels have a very high correlation. This is understandable as the clarification question in more than 95% of the clarification panes in the dataset is the general question of "*Select one to refine your search*". Therefore, the overall quality of a clarification pane is mainly based on the quality of its options. Hence, this study only focuses on *Overall Quality*. We also do not investigate the *Understandability* label in this study. The mean value of *Understandability* across the *MIMICS-Duo* dataset is 4.6 (out of 5), showing that more than 90% of the workers agreed that the clarification panes were highly understandable. Therefore, this characteristic has a minor impact on our evaluations. Figure 2 shows an overview of variables used in this study from the *MIMICS-Duo* dataset.

*3.2.1* **Overall relationship between online and offline evaluations.** The main aim of this research is to compare the clarification ranked lists created using offline labels with the ideal clarification ranked lists created using the *Engagement Level* (i.e., the ground truth), in general, and to compare the top-rated ones in the ranked lists, in particular. Figure 3 shows an example of ranking three clarification panes $[A, B, C]$ for a given query "Gift for grandfather" if the corresponding online *Engagement Levels*, based on CTR and the *Coverage* label, scored by annotators are $[8, 4, 0]$ and $[3, 5, 1]$, respectively. We can see from this example that the offline label, here *Coverage*, was not completely successful in replicating the ideal ranked list, except for the clarification pane *C*.

In this study, we first investigate the relationship between online and offline labels on all 306 queries in the *MIMICS-Duo* dataset in terms of similarity of the ranked-list of clarifications without applying any filtering or grouping on the dataset. In the next step, we investigate if collected offline labels can be used as input features in LTR models to understand whether the combination of offline labels can produce ranked lists of clarification panes more similar to ideal ranked lists, compared to when the ranked lists are created using individual offline labels. To comprehend the interdependency of the offline labels, Tavakoli et al. [61] examined the correlations among offline labels. They discovered that there was only a week correlation between *Coverage* and *Diversity*, while the remaining labels displayed negligible to low correlations. We use four offline labels of *Overall Quality*, *Coverage*, *Diversity*, and *Importance Order*, as well as the number of candidate answers in each clarification pane as input features in the LTR models. The features are linearly normalised based on their min/max values. Considering its different nature, we do not use the *List-wise Preference* label. While other labels offer insights into various aspects of clarification panes, this label is based on the relative rating of all clarification panes generated for a given query. We employ four LTR models, including *Mart*, *RandomForests*, *RankBoost*, *AdaRank* that are implemented in *RankLib* [17]. We also utilise *SVM-rank* [30][3] with a linear kernel. We use 5-fold cross-validation to evaluate our models. In each fold, the dataset is split into training and testing sets by the ratio of 4:1.

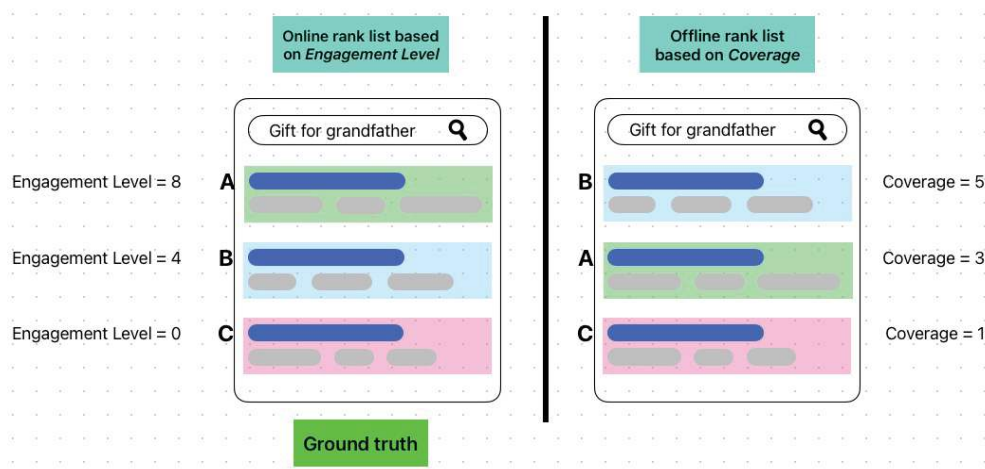

Fig. 3. Two clarification pane rank lists for the query "Gift for grandfather". The left online ranked list is based on the *Engagement Levels* from *Bing* users and acts as our ground truth. The ranked list on the right is an example of an offline rank list based on *Coverage*.

Ultimately, we leverage the potential of GPT-3.5, an advanced Large Language Model, to predict online user engagement with the clarification panes. We use *GPT-3.5-turbo* model.[4] The task assigned to GPT-3.5 is to predict the *Engagement Level* within a range of 0 to 10. Initially, we incorporate the offline labels as input for the model. The prompt that we use to feed the GPT model contains *(1)* a *query*, *(2)* a *clarification pane* that includes *Clarification Question* and associated *Options* (*Candidate Answers*) and *(3)* four *offline labels* similar to LTR models. Subsequently, we conduct the experiment once more, this time excluding the use of offline labels as the model input. This will help determine if the inclusion of offline labels indeed boosts the model's efficacy in predicting user engagement. Our initial experiments explored various prompts that focused on the same task. We noted that when attempting to include offline labels as input, there were cases where GPT-3.5 encountered difficulty in generating the *Engagement Level*. In some instances, it

---

presented the information in a quantitative format rather than within the specified range of 0 to 10. The most successful prompt templates utilised in this study are shown in Figures 4 and 5.[5]

We prompt the model to generate an *Engagement Level* for 1,034 query-clarification pairs. We conduct experiments using various temperature settings, specifically, *temp* = {0.0, 0.5, 1.0}. The temperature parameter regulates the degree of randomness in the generated text. During text generation, the model generates a probability distribution over the next word or token, and the temperature parameter influences the shape of this distribution. A higher temperature value, such as 1.0, results in a more uniform distribution and increases the randomness in the generated output. This can lead to a wider range of diverse and creative responses but may also introduce more errors or nonsensical text. On the other hand, a lower temperature value, such as 0.2, sharpens the distribution, making it narrower and less random. This tends to produce more focused and deterministic responses. Choosing the appropriate temperature value depends on the desired balance between randomness and coherence in the generated text. By experimenting with different *temp* values, we aim to identify the optimal setting for aligning online and offline evaluations in search clarification. Subsequently, we rank the clarification panes for each query based on the predicted *Engagement Level* by GPT-3.5 and compare these rankings against the ideal ranked lists, created using actual *Engagement Level*.

Next, we investigate the impact of query length on the relationships between online and offline evaluations in search clarification. While there is no universal definition of what constitutes a short or long query, some researchers have used a threshold of 3–5 words for short queries and 6 or more words for long queries. For example, Bendersky and Croft [9] defined short queries as those containing up to four words and long queries as those containing five or more words. In another study, Huston and Croft [25] used thresholds of 2, 4, and 5 words to distinguish between very short, short, and long queries. The *MIMICS-Duo* contains queries with a length of 1 to 9 words. However, the number of queries in the dataset for each query length varies. For instance, there are 45 queries with one word, while only 7 queries with 9 words. To investigate the impact of the query length and keep a balance between the groups in terms of the number of queries and query-clarification pair, we assume a query is short if the length is between 1–4 words (126 queries with 415 query-clarification pairs) and it is long if the length is between 5–9 words (180 queries with 619 query-clarification pairs).

### 3.2.2 *Impact of uncertainty in online labelling on corresponding with offline evaluations*. Here, we group the clarification panes based on the *Impression Level* and discard any query-clarification pair with a low *Impression Level*. As mentioned in Section 3.1.1, there is a three-step *Impression Level* per query-clarification pair (i.e., low, medium, high). The *Impression Level* was computed based on the number of times the given query-clarification pair was shown to users. Hence, the *Impression Level* correlates with the query frequency. This highlights the fact that for the query-clarification pairs with low *Impression Level*, the *Engagement Level* obtained by the query-clarification pairs does not necessarily reflect how engaging it was. This part of the study helps us to focus on more reliable data. Removing the query-clarification pairs with the low *Impression Level* leaves the dataset with 212 queries and 703 query-clarification pairs with medium and high *Impression Level* and with one further step of filtering by removing the query-clarification pairs with medium *Impression Level*, 70 queries with 287 query-clarification pairs remain.

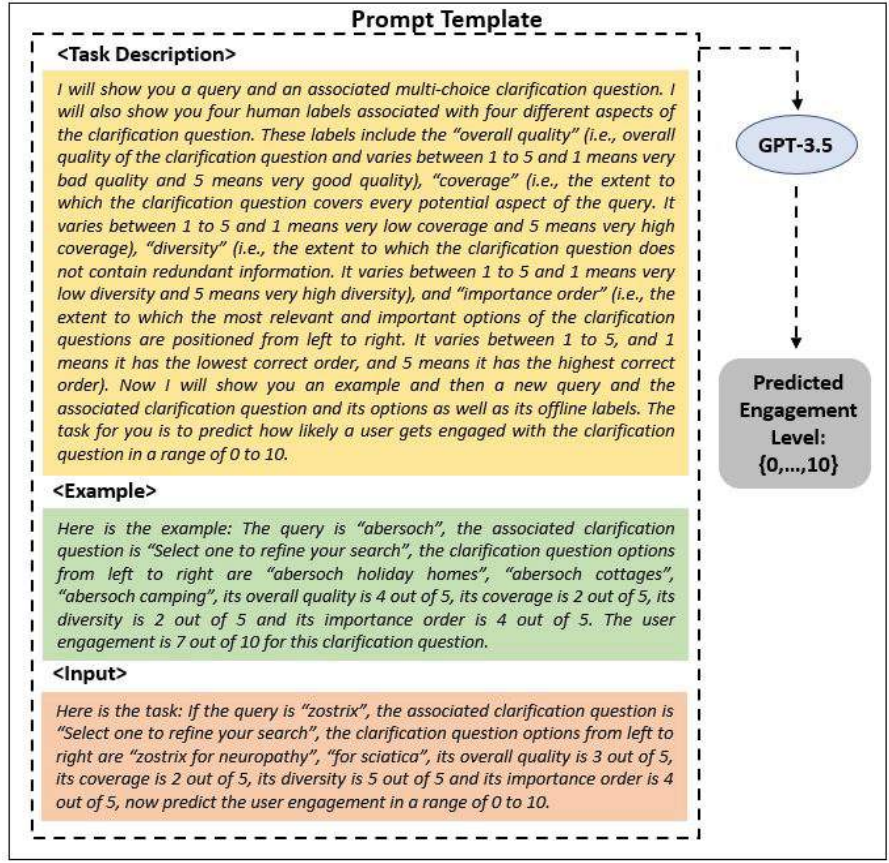

Fig. 4. The prompt template used to feed the GPT model when offline labels were used as the model input.

### 3.3 Evaluation Metrics

As previously stated, this study encompasses two primary objectives: firstly, to assess the effectiveness of offline labels in prioritising the MECP at the top of the list, and secondly, to determine the degree of similarity between the ranked lists produced by the offline labels and the ideal ranked lists for clarification panes. Since the aim of any clarification selection model is to show the MECP to the users (i.e., selecting the most engaging pane among multiple generated clarification panes for a given query), it does not matter whether the clarification pane with the *Engagement Level* of 10 is the top-rated or with the *Engagement Level* of 4. Hence, metrics such as precision at position one (P@1) or mean reciprocal rank (MRR) are appropriate for evaluating the position of the MECP in the ranked list, without taking into account the specific *Engagement Level*. We define P@1 as shown in Eq. 1:

$$P@1 = \frac{TP}{TP + FP} \tag{1}$$

where true positive (TP) and false positive (FP) are the total numbers of clarification panes that are correctly and incorrectly top-rated, respectively, for all queries.

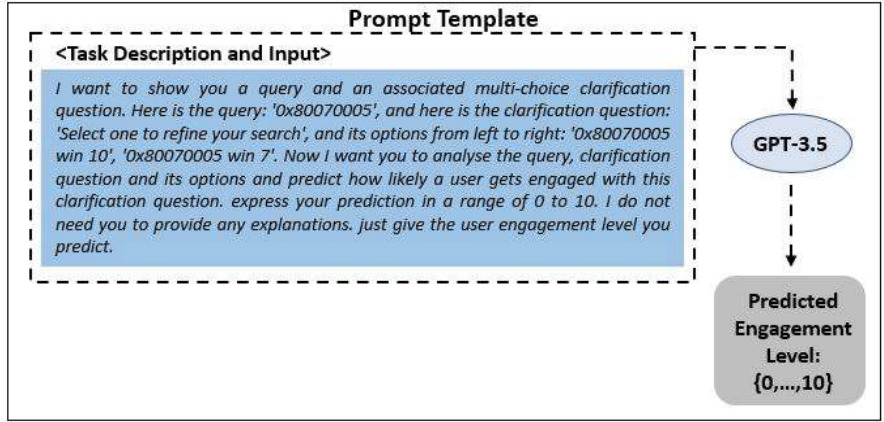

Fig. 5. The prompt template used to feed the GPT model when offline labels were not used as the model input.

To measure MRR, we calculate the reciprocal rank at which the MECP is retrieved in a ranked list of clarification panes and calculate the mean value across all queries. We also measure normalised discounted cumulative gain at position one (NDCG@1) that considers the relevance factor (here, the *Engagement Level*) when evaluating the top-rated clarification pane.

For the second objective, which involves assessing the similarity between the clarification ranked lists, we use NDCG@3. The choice of a cutoff at 3 is based on the observation that approximately 70% of queries consist of only three clarification panes. Furthermore, for queries with four or more clarification panes, around 50% of those panes receive no user engagement. Hence, NDCG@3 ensures a fair evaluation of all clarification panes at a consistent depth.

We also calculate rank-biased precision (RBP) [46] and ranked-biased overlap (RBO) [67] that consider a binary relevance factor in the evaluation of the top-rated clarification pane in the list. RBP measures the utility rate that is gained by a user at a given degree of persistence ($p$), representing an aspect of user behaviour. Moffat and Zobel [46] assumed that a user inspects the first document and proceeds from the *ith* document to the *i+1th* with fixed conditional probability $p$. For instance, if $p$=0.5, the user obtains a high average per document utility, which means there is a relevant document in the first one or two rank positions. The RBP equation (Eq. 2) is proposed below:

$$RBP = (1-p)\sum_{i=1}^{d} r_i \cdot p^{i-1} \tag{2}$$

where $r_i$ indicates the binary relevance of the *ith* ranked document scored as either 0 (not relevant) or 1 (relevant).

The RBP metric was introduced to measure the effectiveness of a ranked list retrieved for a query and varies between 0 and 1. However, RBP cannot be used directly in this study as only one clarification pane is shown to a user at a time, not a list of clarification panes. To employ RBP in this study, we assume: *(1)* regardless of the value of *Engagement Level*, if there is a positive *Engagement Level* for a given clarification pane, $r_i$=1 and if not, $r_i$=0, and *(2)* since only one pane is shown to a user, we assume $p$=0.05, which means the probability of a user checking the second clarification pane (if it exists) is roughly 5%. We also calculate RBP for $p$ values of 0.5 and 0.7 to investigate the clarification pane ranked lists at deeper depths. We calculate RBP for every ranked list generated by each offline label and report the average RBP for each label.

Table 1. Relationships between the ranked lists of clarification panes created by the *Engagement Level* and created by offline labels.

| Engagement Level vs. | Metric | | | | | |
|---|---|---|---|---|---|---|
| | **NDCG@1** | **NDCG@3** | **P@1** | **MRR** | **RBP** | **RBO** |
| *List-wise Preference* | **0.459** | 0.729 | 0.559† | 0.749† | **0.520** | **0.339** |
| Aspect *Overall Quality* | 0.433 | 0.724 | 0.562† | **0.760**† | 0.503 | 0.301 |
| *Coverage* | 0.448 | 0.725 | **0.569**† | 0.747† | 0.510 | 0.329 |
| *Diversity* | 0.454 | **0.731** | 0.523† | 0.726† | 0.515 | 0.323 |
| *Importance Order* | 0.412 | 0.706 | 0.484† | 0.710† | 0.455 | 0.275 |
| *Mean* | 0.438 | 0.723 | 0.535 | 0.736 | 0.496 | 0.307 |
| *Random Ranker* | 0.403 | 0.706 | 0.307 | 0.561 | 0.469 | 0.285 |

† Significantly different from the Random Ranker baseline (Tukey HSD test, *p<0.05*).

The second rank-biased metric is RBO, developed by Webber et al. [67] and is a similarity measure to compare two ranked lists, quantifying how far the observed ranking deviates from the ideal ranking. It has the same assumptions as RBP and can be calculated using the Eq. 3:

$$RBO = (1 - p) \sum_{k=1}^{\infty} p^{k-1} \frac{\left| A_{1:k} \cap B_{1:k} \right|}{k} \qquad (3)$$

where $A$ and $B$ are two ranked lists, $k$ is the depth of comparison, $\left| A_{1:k} \cap B_{1:k} \right|$ is the size of intersection between two lists at depth $k$.

RBO varies between 0 and 1; 1 means both ranked lists are identical, and 0 means they are completely disjoint. It is evident that RBO investigates the overlap and ordering between two ranked lists (the number of identical documents shared between two ranked lists). The current RBO definition cannot be used in this study as the clarification panes for a given query in the ranked lists generated by any two labels are always the same. Therefore, RBO in the current definition is always 1. To adopt RBO in this study, we define the size of the intersection of two ranked lists based on the number of panes that have the same positions in both lists. We calculate RBO between the ideal ranked list generated by *Engagement Level* and ranked lists generated by offline labels.

## 4 RESULTS

We present the results of experiments on online–offline evaluations in search clarification in the following subsections.

### 4.1 Overall Relationship Between Online and Offline Evaluations

First, the offline labels were used individually to create the clarification ranked lists and then the offline labels were employed as input features for LTR and GPT-3.5 models to create the ranked lists. In the following step, we repeated the experiments on the short and long queries. To assess the performance of the offline labels in comparison to a baseline, we additionally ranked the clarification panes for each query using a Random Ranker.[6] For the sake of reproducibility, our results and codes are publicly available.[7] We performed Tukey honestly significant difference (HSD) [64] to find the means that were significantly different from each other for each column in the tables. The Tukey HSD test is a post hoc test used when there are equal numbers of subjects in each group for which pairwise comparisons of the data are made [59]. The highest-performing label is highlighted in bold within each column in all presented tables.

---

[6]Random Ranker is repeated 1000 times, and the mean values are reported.
[7]https://github.com/Leila-Ta/On_Off-Eval-Search_Clarification

*4.1.1* **Offline labels.** Table 1 shows the relationships between the ranked lists of clarification panes created by the Engagement Level and the ranked lists created by offline labels on all queries. We can observe that *(1)* the MECPs were more likely to have the highest *Overall Quality* and *Coverage* compared to other clarification panes; *(2)* all offline labels performed noticeably better than a Random Ranker (e.g., *Coverage* showed 85% improvement over a Random Ranker in presenting the MECP for a given query at the top of the ranked list). However, *Importance Order* evaluation methodology showed the poorest performance among all offline methods. These findings were derived from the P@1 and MRR metrics analysis, revealing statistically significant differences between them. The slight improvements over a Random Ranker shown by other metrics (i.e., NDCG@1, NDCG@3, RBP, and RBO) were not significant. This indicates that the metrics used to compare online and offline evaluations in search clarification have noticeable influences on the result justifications. For instance, P@1 and MRR are unconcerned about the user *Engagement Level* and they only check the rank of the MECP. While for NDCG@1, if an engaging clarification that is not the MECP is ranked top, it still receives a score. Such an evaluation increases the chance of a Random Ranker showing a better performance than when the evaluation is only based on the position of the MECP. As indicated in Section 3.3, we also calculated RBP and RBO for two higher $p$ values (i.e., 0.5 and 0.7) in addition to 0.05 that are shown in Table 1 to investigate the similarity in the ranked lists at deeper depths. We observed that the performance of offline labels merged toward a Random Ranker by increasing the $p$ value (see Figure 6).

We also considered the Kendall ($\tau$) [35], and Spearman ($r_s$) [68] rank correlations between online and offline ranked lists generated for each query but did not observe correlations. The majority (70%) of the ranked lists only had three clarification panes, and such a correlation analysis may not be accurate enough to draw conclusions. However, a less sensitive analysis using Pearson correlations [50] across all query-clarification pairs captured weak correlations between two offline labels of *Overall Quality* and *List-wise Preference* with the *Engagement Level* (i.e., $\rho$=0.304 between *Overall Quality* and *Engagement Level* and $\rho$=0.316 between *List-wise Preference* and *Engagement Level*).

*4.1.2* **LTR models.** During the second phase of the experiments, our objective was to investigate how the combinations of offline labels impact the relationship between online and offline evaluations. We formulated this experiment as an LTR task and incorporated the offline labels as input features for the models. The performances of the LTR in ranking the clarification panes are shown in Table 2. It is evident that *SVM-rank* exhibited better performance compared to other LTR models. However, its superior performance was not significantly different from the other LTR models. When evaluating the effectiveness of LTR models using P@1 and MRR and comparing them to the *Overall Quality* or *Coverage* labels in Table 1 (two outperforming offline labels based on the same metrics), it becomes apparent that LTR models that incorporated the offline labels as input features did not outperform the individual offline labels in accurately ranking the MECPs at the highest position in the lists. However, the performances of *SVM-rank* and *AdaRank* were significantly better than the Random Ranker, presented in Table 1. It seems the complexity of the LTR models may not be adequate to capture the underlying patterns present in the data. Furthermore, the characteristics and size of the training data can also impact the performance of LTR models, posing a challenge for the models to effectively learn robust patterns and generalise effectively.

*4.1.3* **Large language model.** Table 2 also indicates the performance of GPT-3.5 in predicting user engagement and ranking clarification panes. We examined GPT-3.5 using three different temperature settings: 0.0, 0.5, and 1.0. Comparing Table 1 and 2 reveals that not only GPT-3.5 outperformed LTR models in terms of P@1 and MRR when a temperature of 0.0, 0.5 and 1.0 are utilized, but it also showed significantly better performance compared to the individual offline labels of *Overall Quality* and *Coverage* when a temperature of 0.0 is used. Obtaining the best results with a temperature value of 0 suggests that GPT-3.5 has achieved optimal performance by using a deterministic approach. This deterministic behaviour is advantageous when we want to prioritise consistency and precise predictions. However, it is important to note that using a temperature of 0 may lead to overly rigid and repetitive outputs, as the lack of randomness can result in a lack of diversity. When the temperature value is set to 0, it means that the output generated by GPT-3.5 is determined solely by the model's confidence scores. In other words, the model selects the most probable word or token at each step without any randomness or variation. This finding emphasises the efficacy of GPT-3.5 in predicting online user engagement and hence, accurately identifying the MECPs when incorporating the offline labels as the model input. However, similar to LTR models and offline labels, GPT-3.5 fell short of significantly surpassing the performance of the Random Ranker in ranking multiple clarification panes for given queries (no significant differences were observed between the performances of GPT-3.5 and the Random Ranker in terms of NDCG@3.

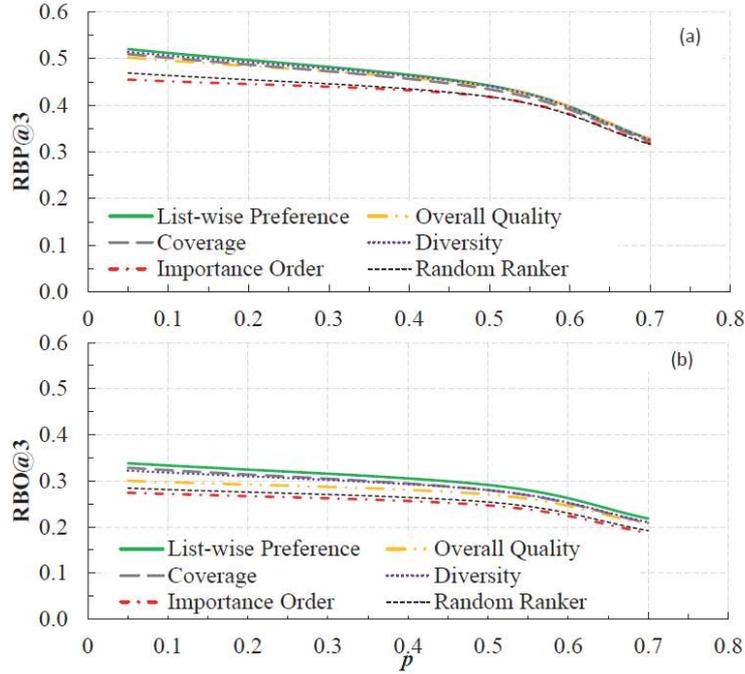

Fig. 6. Variations of (a) RBP and (b) RBO at a depth of 3 for different values of *p*.

We also observed that when GPT-3.5 was provided with high-quality human-annotated labels of clarification characteristics, it showed better performance compared to the *List-wise Preference* labelling approach conducted by crowd-source workers. In the crowd-sourcing task, all the generated clarification panes for a given query were presented to workers simultaneously, and the workers were asked to rate all the panes based on their preferences (without having access to the *Aspect* labels). Although GPT-3.5 could not predict the relative *Engagement Level* among the panes and evaluated each pane independently, its user engagement prediction resulted in more successful identification of the MECPs compared to the *List-wise Preference* labelling method.

*4.1.4  **Impact of query length on the relationship between online and offline evaluations.*** Table 3 shows the calculated metrics for short (1–4 words) and long (5–9 words) queries. If a query is short, the *List-wise Preference* evaluation performs better than other offline labels in placing the MECP at rank one (i.e., obtaining the highest P@1, MRR and RBO). However, if the query is long, selecting the MECP from a pool of clarification panes generated for a query can be carried out using *Overall Quality* and *Coverage* evaluations. Similar to the previous table, no conclusion can be drawn about the impact of the query length on the similarity of the ranked lists, as they did not show any significant improvement over a Random Ranker (no significant differences were measured in NDCG@3 between offline labels and the Random Ranker). We also performed a Tukey HSD test on the calculated P@1 and MRR values for short, long, and all queries. The results indicate that there are no significant differences, suggesting that the length of the query does not have an impact on the relationship between offline evaluations and online evaluations in the context of search clarification.

Table 2. Evaluation of three GPT-3.5 configurations across varying temperature settings and five LTR models, utilising offline labels to generate ranked lists of clarifications.

| Engagement Level vs. | Metric | | | | | |
|---|---|---|---|---|---|---|
| | NDCG @1 | NDCG @3 | P@1 | MRR | RBP | RBO |
| *RandomForests* | **0.473** | 0.739 | 0.357$^{†‡£}$ | 0.611$^{†‡£}$ | 0.507 | **0.358** |
| *AdaRank* | 0.472 | 0.736 | 0.426$^{†‡£§}$ | 0.673$^{†‡£§}$ | 0.498 | 0.340 |
| *MART* | 0.468 | 0.733 | 0.341$^{†‡£}$ | 0.609$^{†‡£}$ | **0.508** | 0.342 |
| *RankBoost* | 0.459 | 0.733 | 0.364$^{†‡£}$ | 0.639$^{†‡£}$ | 0.486 | 0.345 |
| *SVM-rank* | 0.456 | **0.741** | **0.427$^{†‡£§}$** | **0.698$^{†‡£§}$** | 0.495 | 0.346 |
| *GPT-3.5* (*temp* = 0.0) | 0.460 | **0.734** | **0.663$^{†§*}$** | **0.830$^{†§\$}$** | **0.525** | 0.382 |
| *GPT-3.5* (*temp* = 0.5) | 0.439 | 0.718 | 0.588$^{§}$ | 0.778$^{§}$ | 0.487 | 0.363 |
| *GPT-3.5* (*temp* = 1.0) | **0.468** | 0.732 | 0.539$^{§}$ | 0.751$^{§}$ | 0.523 | **0.386** |

$^{†, ‡, £}$ Significantly different from GPT-3.5 with *temp* = 1.0, *temp* = 0.5, and *temp* = 0.0, respectively.
$^{§}$ Significantly different from the Random Ranker baseline (Table 1).
$^{*}$ Significantly different from *Coverage*, the best performing label in terms of P@1, Table 1.
$^{\$}$ Significantly different from *Overall Quality*, the best performing label in terms of MRR, Table 1.

### 4.2 Impact of Uncertainty on the Relationship Between Online and Offline Evaluations

Here, we separated the query-clarification pairs based on the *Impression Level* and repeated the experiments (i.e., assessing the position of the MECPs in the created ranked lists and the similarity of the ranked lists). We learned from Zamani et al. [74] that a clarification pane with high *Impression Level* was shown to the users more than a clarification pane with low *Impression Level*. Therefore, the obtained *Engagement Level* by a clarification pane with a high *Impression Level* is likely to be more reliable. In other words, the uncertainty in the collected online data is less. Table 4 shows the calculated metrics for all offline labels for the query-clarification pairs with high *Impression Level* (top section) and with medium and high *Impression Levels* (bottom section). Table 4 indicates that when query-clarification pairs with low *Impression Level* were removed from the dataset (i.e., eliminating uncertainty from online evaluation), the clarification panes with the highest *Overall Quality* were likely to be the MECPs (obtaining high values of P@1 and MRR). However, no significant differences over a Random Ranker were observed for NDCG@3, showing that the offline labels were unable to produce clarification ranked lists better than a Random Ranker.

By simultaneously examining Tables 1, 3, and 4, it becomes evident that the *Importance Order* had the poorest relationship with the online label compared to other offline labels. This implies that the engagement of users with the clarification pane was not significantly influenced by the order of candidate answers. Moreover, comparing Tables 1 and 4 shows much higher values for P@1 and MRR when we removed the query-clarification pairs with low *Impression Level* from the dataset. We performed a Tukey HSD test on the calculated P@1 and MRR values for *Overall Quality* between high *Impression Level* query-clarification pairs (top section in Tables 4) and all query-clarification pairs (Table 1) and between medium and high *Impression Level* query-clarification pairs (bottom section in Tables 4) and all query-clarification pairs (Table 1). The results indicated a significant difference between the two. This suggests that offline evaluation aligned more closely with online evaluation when the uncertainty in online evaluation was minimal, and the observed differences were unlikely to be random occurrences due to the sample size.

Table 3. Impact of the query length on relationships between the ranked lists of clarifications created by the *Engagement Level* and created by offline labels. (Short Query: 126 queries with 415 query-clarification pairs; Long Query: 180 queries with 619 query-clarification pairs.)

| **Engagement Level vs.** | | | **Metric** | | | | | |
|---|---|---|---|---|---|---|---|---|
| | | | NDCG @1 | NDCG @3 | P@1 | MRR | RBP | RBO |
| **Short Query (1–4)** | *List-wise Preference* | | **0.461** | 0.721 | **0.561**$^\dagger$ | **0.751**$^\dagger$ | 0.495 | **0.368**$^\dagger$ |
| | Aspect | *Overall Quality* | 0.408 | 0.707 | 0.539$^\dagger$ | 0.748$^\dagger$ | 0.495 | 0.280 |
| | | *Coverage* | 0.412 | 0.702 | 0.539$^\dagger$ | 0.737$^\dagger$ | 0.473 | 0.317 |
| | | *Diversity* | 0.455 | **0.725** | 0.533$^\dagger$ | 0.737$^\dagger$ | **0.511** | 0.362$^\dagger$ |
| | | *Importance Order* | 0.371 | 0.680 | 0.478$^\dagger$ | 0.710$^\dagger$ | 0.422 | 0.269 |
| | | *Mean* | 0.412 | 0.704 | 0.522 | 0.733 | 0.475 | 0.307 |
| | *Random Ranker* | | 0.376 | 0.684 | 0.289 | 0.550 | 0.422 | 0.259 |
| **Long Query (5–9)** | *List-wise Preference* | | 0.458 | 0.740 | 0.556$^\dagger$ | 0.745$^\dagger$ | 0.549 | 0.300 |
| | Aspect | *Overall Quality* | 0.469 | 0.748 | 0.595$^\dagger$ | **0.777**$^\dagger$ | 0.490 | 0.325 |
| | | *Coverage* | **0.498** | **0.758** | **0.611**$^\dagger$ | 0.762$^\dagger$ | **0.554** | **0.348** |
| | | *Diversity* | 0.452 | 0.741 | 0.508$^\dagger$ | 0.712$^\dagger$ | 0.512 | 0.270 |
| | | *Importance Order* | 0.472 | 0.743 | 0.492$^\dagger$ | 0.710$^\dagger$ | 0.503 | 0.293 |
| | | *Mean* | 0.473 | 0.748 | 0.552 | 0.740 | 0.515 | 0.309 |
| | *Random Ranker* | | 0.441 | 0.739 | 0.333 | 0.578 | 0.516 | 0.302 |

$^\dagger$ Significantly different from the Random Ranker baseline (Tukey HSD test, *p<0.05*).

Table 4. Impact of the *Impression Level* on relationships between the ranked lists of clarifications created by the *Engagement Level* and created by offline labels.

| **Engagement Level vs.** | | | **Metric** | | | | | |
|---|---|---|---|---|---|---|---|---|
| | | | NDCG @1 | NDCG @3 | P@1 | MRR | RBP | RBO |
| **High** | *List-wise Preference* | | 0.617 | 0.837 | 0.614$^\dagger$ | 0.781$^\dagger$ | 0.701 | 0.417 |
| | Aspect | *Overall Quality* | **0.667** | **0.860** | **0.729**$^{\dagger\S}$ | **0.848**$^{\dagger\S}$ | **0.793** | **0.475** |
| | | *Coverage* | 0.657 | 0.849 | 0.657$^\dagger$ | 0.785$^\dagger$ | 0.765 | 0.461 |
| | | *Diversity* | 0.649 | 0.842 | 0.649$^\dagger$ | 0.782$^\dagger$ | 0.740 | 0.449 |
| | | *Importance Order* | 0.577 | 0.818 | 0.614$^\dagger$ | 0.764$^\dagger$ | 0.714 | 0.305 |
| | | *Mean* | 0.638 | 0.842 | 0.661 | 0.795 | 0.753 | 0.423 |
| | *Random Ranker* | | 0.626 | 0.841 | 0.429 | 0.644 | 0.751 | 0.360 |
| **Medium–High** | *List-wise Preference* | | 0.524 | 0.765 | 0.623$^\dagger$ | 0.789$^\dagger$ | 0.588 | **0.427** |
| | Aspect | *Overall Quality* | 0.533 | **0.776** | **0.665**$^{\dagger\S}$ | **0.816**$^{\dagger\S}$ | 0.606 | 0.405 |
| | | *Coverage* | **0.535** | 0.772 | 0.618$^\dagger$ | 0.775$^\dagger$ | **0.613** | 0.404 |
| | | *Diversity* | 0.528 | 0.772 | 0.613$^\dagger$ | 0.773$^\dagger$ | 0.597 | 0.409 |
| | | *Importance Order* | 0.446 | 0.734 | 0.519$^\dagger$ | 0.731$^\dagger$ | 0.499 | 0.303 |
| | | *Mean* | 0.511 | 0.764 | 0.604 | 0.774 | 0.579 | 0.380 |
| | *Random Ranker* | | 0.473 | 0.744 | 0.401 | 0.634 | 0.553 | 0.357 |

$^\dagger$ Significantly different from the Random Ranker baseline.
$^\S$ Significantly different from the same metric calculated on all query-clarification pairs in Table 1.

Table 5. Impact of the *Impression Level* on the performance of three GPT-3.5 configurations across varying temperature settings.)

| Impression Level | Engagement Level vs. | Metric | | | |
|---|---|---|---|---|---|
| | | NDCG @1 | NDCG @3 | P@1 | MRR |
| High | GPT-3.5 (*temp* = 0.0) | **0.658**$^\dagger$ | **0.860**$^\dagger$ | **0.786**$^\dagger$ | **0.890**$^\dagger$ |
| | GPT-3.5 (*temp* = 0.5) | 0.648$^\dagger$ | 0.844$^\dagger$ | 0.657 | 0.821 |
| | GPT-3.5 (*temp* = 1.0) | 0.614$^\dagger$ | 0.828$^\dagger$ | 0.529 | 0.749 |
| Low−Med.−High | GPT-3.5 (*temp* = 0.0) | 0.460 | **0.734** | **0.663** | **0.830** |
| | GPT-3.5 (*temp* = 0.5) | 0.439 | 0.718 | 0.588 | 0.778 |
| | GPT-3.5 (*temp* = 1.0) | **0.468** | 0.732 | 0.539 | 0.751 |

$^\dagger$ Significantly different from GPT-3.5 with the same *temp* when using all query-clarifictaion pairs.

Additionally, we conducted GPT prompts using query-clarification pairs that only had a high *Impression Level* (top section in Table 5). We then compared the model's performance in predicting the *Engagement Level* with the results obtained when using all query-clarification pairs (bottom section in Table 5). We only measured P@1, MRR, NDCG@1 and NDCG@3 here as the metrics of RBP and RBO did not show the required capabilities for such comparisons. The results indicated a significant improvement in GPT-3.5 performance, particularly when using *temp* = 0.0, compared to when using all query-clarification pairs. According to the findings presented in Table 5, when there is reduced uncertainty in the online evaluation, the performance of GPT-3.5 in predicting online user engagement improves when the GPT prompt includes offline labels.

In the final phase of comprehending the relationship between online and offline assessments in search clarification, we employed GPT-3.5 to predict the *Engagement Level* without using offline labels as input for the model. We conducted this experiment initially on all 1,034 query-clarification pairs, and subsequently on 287 pairs with a high *Impression Level*. Tables 6 and 7 showcase GPT's performance in predicting the *Engagement Level*, both with and without the incorporation of offline labels as model inputs. It is evident that integrating offline labels as input for GPT-3.5 enhances its capacity to anticipate user engagement. Despite outperforming individual offline labels and LTR models in predicting user engagement when integrated with offline labels, GPT's performance notably declined in identifying the MECPs and generating ranked lists of clarification similar to ideal ranked lists when used independently (not using offline labels as the model input). Surprisingly, it even demonstrated lower effectiveness compared to certain offline labels. This observation underscores the significance of offline labels in predicting online user engagement, emphasising that despite the recent enhancement in language models, they still cannot entirely replace human assessments in every situation.

Table 6. Impact of offline labels on the performance of three GPT-3.5 configurations across varying temperature settings on the entire dataset.

| Model Input | Engagement Level vs. | Metric | | | |
|---|---|---|---|---|---|
| | | NDCG @1 | NDCG @3 | P@1 | MRR |
| Using Offline Labels | *GPT-3.5* (*temp* = 0.0) | 0.460† | **0.734**† | **0.663**† | **0.830**† |
| | *GPT-3.5* (*temp* = 0.5) | 0.439† | 0.718† | 0.588† | 0.778† |
| | *GPT-3.5* (*temp* = 1.0) | **0.468**† | 0.732† | 0.539† | 0.751† |
| Not Using Offline Labels | *GPT-3.5* (*temp* = 0.0) | 0.346 | 0.626 | **0.587** | **0.703** |
| | *GPT-3.5* (*temp* = 0.5) | 0.338 | 0.618 | 0.448 | 0.607 |
| | *GPT-3.5* (*temp* = 1.0) | **0.390** | **0.649** | 0.340 | 0.623 |

† Significantly different from GPT-3.5 with the same *temp* but without using offline labels as input for the model.

Table 7. Impact of offline labels on the performance of three GPT-3.5 configurations across varying temperature settings on only query-clarification pairs with High *Impression Level*.

| Model Input | Engagement Level vs. | Metric | | | |
|---|---|---|---|---|---|
| | | NDCG @1 | NDCG @3 | P@1 | MRR |
| Using Offline Labels | *GPT-3.5* (*temp* = 0.0) | **0.658** | **0.860** | **0.786**† | **0.890**† |
| | *GPT-3.5* (*temp* = 0.5) | 0.648† | 0.844 | 0.657† | 0.821† |
| | *GPT-3.5* (*temp* = 1.0) | 0.614 | 0.828 | 0.529† | 0.749† |
| Not Using Offline Labels | *GPT-3.5* (*temp* = 0.0) | 0.609 | 0.825 | **0.600** | **0.720** |
| | *GPT-3.5* (*temp* = 0.5) | 0.552 | 0.816 | 0.404 | 0.717 |
| | *GPT-3.5* (*temp* = 1.0) | **0.621** | **0.837** | 0.404 | 0.615 |

† Significantly different from GPT-3.5 with the same *temp* but without using offline labels as input for the model.

### 4.3 The Most vs. the Least Engaging Panes

To enhance our understanding of how the offline labels correspond with the online label in MECPs, we compared the most engaging clarification panes with the least engaging clarification panes (LECPs) for queries that their clarification panes had high *Impression Level*. High *Impression Level* query-clarification pairs were chosen to ensure that the uncertainty in the low *Engagement Level* obtained by the LECPs is minimal. We observed that the *Overall Quality* of MECPs was higher than of the LECPs for more than 51% of the MECPs and it agrees with our observations in Table 4 (see Figure 7). Although the percentage of the MECPs with higher *Coverage*, *Diversity* and the number of candidate answers were also higher than the LECPs, but the observed higher percentages were not significantly different according to Student's t-test. This indicates the *Overall Quality* of a clarification pane contributed to making it engaging from a user's perspective.

### 4.4 Manual Clarification Pane Inspection

To explore the scenarios where a clarification pane with low quality might engage users more than a high-quality pane, we conducted a manual inspection of two queries. For these queries, the online and offline labels did not align well with their MECPs and LECPs. The details of this analysis can be found in Tables 8 and 9.

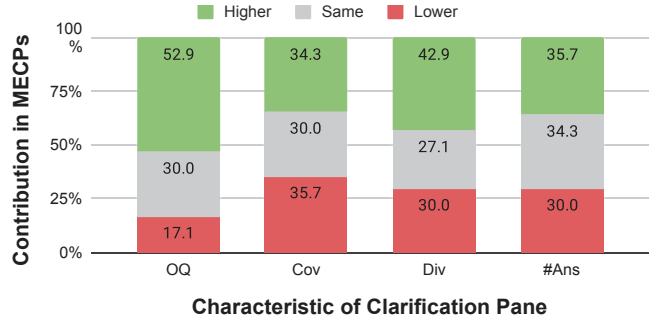

**Fig. 7.** Variations of *Overall Quality* (OQ), *Coverage* (Cov), *Diversity* (Div) and the number of candidate answers (# Ans) in the MECPs when compared to the LECPs.

Table 8. Examples queries and their most and least engaging clarification panes.

| Query | Pane | Clarification Options | | | | |
|---|---|---|---|---|---|---|
| | | Option 1 | Option 2 | Option 3 | Option 4 | Option 5 |
| **yucca** | MECP | yucca valley | yucca mountain | yucca desert | yucca lake | yucca canyon |
| | LECP | yucca benefits | yucca nutrition facts | yucca powder | yucca for sale | *null* |
| **why is my printer offline** | MECP | hp | why is my printer offline dell | *null* | *null* | *null* |
| | LECP | in windows 10 | windows 8 | windows 7 | windows xp | *null* |

In the case of the first query, "*yucca*", the term can potentially refer to either a shrub or Yucca Mountain in Nevada, USA. The MECP is associated with the mountain, whereas the LECP is related to the plant. Upon analysing the clarification options for the MECP, we observed that they predominantly focused on a single intent and exhibited limited diversity. Specifically, terms such as "mountain", "valley", and "canyon" represented similar aspects of Yucca Mountain. Conversely, the clarification options for the LECP encompassed aspects of the yucca plant, indicating a greater diversity in the coverage of relevant information (see Tables 8).

According to Tavakoli et al. [61], in the data collection process, the workers were initially presented with the query and eight associated retrieved documents before annotating a label. Each retrieved document included a title and snippet. The workers were instructed to review these documents to understand various aspects related to the query before proceeding with the labelling task. In the case of the "yucca" query, we noticed that all the retrieved documents shown to the workers focused on the shrub, with no documents about the mountain. It is speculated that the workers inferred the query's intent based on the content they reviewed in these documents and performed the labelling task with that intent in mind. However, the users recorded in the online data got more engaged with a different clarification pane, which covered the query's intent not reflected in the retrieved documents (see Table 9). This suggests that as long as a clarification pane addresses an aspect of the query that is absent in the retrieved documents, users are likely to engage with it, irrespective of its quality.

For the second query, "*why is my printer offline*", the MECP asked for the printer brand, while the LECP requested clarity from a software point of view. The coverage and diversity labels for both clarification panes were shallow and correctly rated by the human annotators. However, the annotators believed that the LECP had higher quality than the MECP as it perhaps provided more options than the MECP, with only two options. Upon reviewing the retrieved documents, it becomes evident again that all of them are focused on printer issues occurring on various versions of Windows. None of the documents provide information specifically related to the brand of the printer.

Table 9. Examples queries with online and offline labels.

| Query | Pane | Engagement Level | Overall Quality | Coverage | Diversity |
|---|---|---|---|---|---|
| **yucca** | MECP | 3 | 4 | 4 | 3 |
| | LECP | 1 | 5 | 3 | 4 |
| **why is my printer offline** | MECP | 8 | 3 | 2 | 2 |
| | LECP | 0 | 5 | 2 | 3 |

Examining these two examples underscores the significance of soliciting clarification questions from users when the quality of retrieved documents is subpar. Moreover, it reveals that the accuracy of offline labelling is greatly influenced by the information provided to the workers before the labelling process and their knowledge about the query in some instances.

## 5 DISCUSSION

We showed that the evaluation of retrieval quality through online and offline assessments often produces contrasting results, as observed in previous studies on this topic [16, 18, 22, 56]. Specifically, the findings of our current research differ from those of a prior study focused on search clarification [74]. Zamani et al. [74] examined the *MIMICS* dataset and investigated correlations between online and offline evaluations using a single offline label. They concluded that no correlation was observed between the two evaluation methods. In contrast, the current study analysed the *MIMICS-Duo* dataset utilising various online and offline labels. We observed a relationship between online and offline evaluations in the context of search clarification when the aim is to identify the most engaging clarification pane among multiple generated panes for a given query. However, our research supports previous studies by revealing a discrepancy between online and offline evaluations regarding ranking clarification panes for a given query.

We manually examined various panes to understand why users might engage more with lower-quality clarification panes. We observed that while the human annotation was carried out accurately based on the available information, it does not always guarantee that the annotators can accurately capture the user's intent. This finding helps to explain the contradictions observed between online and offline evaluations.

In attempting to explain these discrepancies, we consider two explanations proposed by Teevan et al. [63] and Liu et al. [44]. Teevan et al. [63] suggested that different users who issue the same textual query may have distinct information needs or intentions, leading to varying evaluations. This implies that users' subjective preferences and expectations play a significant role in assessing the quality of clarification panes. Liu et al. [44], on the other hand, proposed that there may be notable disparities between assessors' judgements and users' assessments due to differences between satisfaction prediction and document relevancy prediction. To some extent, satisfaction is subjective, as different users may have varying opinions on what constitutes a satisfying experience.

Apart from the reasons mentioned here, it is essential to acknowledge that the information provided to annotators can impact the correlation between online and offline evaluations. When determining the MECPs, it is essential to assess the SERP and clarification pane quality as well as their relation to each other. Evaluating either component independently may lead to misleading conclusions in certain scenarios.

This study demonstrates the value of using collected offline labels for predicting online user behaviour and identifying the MECP within generated panes for a query, particularly when employing Language Models for task formulation. Despite having identical input features, we observed different performances between the GPT-3.5 and LTR models. The observations can be attributed to several factors:

- Model Complexity and Training Data: GPT-3.5 is a highly complex language model with 175 billion parameters. It has been trained on a large and diverse corpus of text from the internet, which gives it a broad understanding of natural language. This extensive training data allows it to make nuanced judgements about relevance [19]. However, the LTR model had no access to such a vast and diverse dataset. Moreover, The LTR model might have been trained on a dataset that introduced some biases or limitations that affected its performance. GPT-3.5's extensive pre-training on diverse internet text might have helped it overcome some of these biases.
- Contextual Understanding: GPT-3.5, with its deep transformer architecture, has been trained to generate human-like text based on context. It can learn from vast amounts of data and this context awareness might enable it to better understand the relationship between queries, clarification questions, options, relevance labels and user engagement.
- Model Architecture: GPT-3.5 and LTR models have different architectures and underlying principles. GPT-3.5 is a transformer-based language model that excels at capturing semantic and contextual information in text. On the other hand, LTR models, such as AdaRank or RankBoost, are specifically designed for learning to rank tasks and may have different assumptions and optimisations.
- Learning Approaches: GPT-3.5 utilises unsupervised learning through language modelling objectives, which allows it to capture a wide range of language patterns and contexts. In contrast, LTR models often rely on supervised learning techniques with explicit relevance labels or features specific to ranking tasks.
- Evaluation Metric: The metric used to evaluate performance might favour GPT-3.5's capabilities. If the task relies heavily on natural language understanding and generation, GPT-3.5's strengths would be more pronounced.
- Generalisation Ability: GPT-3.5 is designed to generalise well across a wide range of tasks without task-specific fine-tuning. This means it can handle a diverse set of queries and situations effectively, including those it wasn't explicitly trained for.

The observations and findings in this research have several theoretical and practical implications as following:

- By investigating the relationship between online and offline evaluations specifically in the context of search clarification, we contribute to a deeper theoretical understanding of how offline assessments relate to real-time user engagement.
- By understanding which characteristics contribute most to engagement, developers can tailor their approaches to better meet user needs and preferences.
- Insights from our study can inform the development of evaluation methods for search systems. By considering both online and offline evaluation approaches and understanding their relationship, researchers and practitioners can design more comprehensive evaluation frameworks that capture the nuanced aspects of user engagement.
- The finding that Large Language Models outperform Learning-to-Rank models and individual offline labels suggests practical implications for model selection and integration in search systems. Integrating human labels into model training can enhance the performance of LLMs, leading to more accurate identification of engaging clarification questions from the user's perspective.

## 6  CONCLUSIONS AND FUTURE WORK

How well online and offline evaluations correspond to each other in search clarification is the knowledge gap that was addressed in this study by answering the research questions below:

**RQ1**: How well do offline evaluations correspond with online evaluations in search clarification?

Offline evaluations can complement online evaluations in identifying the most engaging clarification pane for a given query. This suggests that offline evaluation methodologies can be useful for assessing the effectiveness of search clarification models in terms of user engagement. We have demonstrated that clarification panes must excel in multiple aspects to be considered engaging from a user's perspective. Merely having high *Coverage* or *Diversity* does not guarantee engagement. However, when ranking multiple clarification panes for a given query, offline evaluations do not outperform a Random Ranker. This implies that current offline evaluation methodologies may not be well-suited for evaluating the ranking performance of search clarification models. We also showed that some offline labels, in particular, *Overall Quality* and *Coverage* perform better than others in corresponding with user engagement.

We automated the ranking of clarification panes to identify the MECP from a user's perspective for a given query using GPT-3.5 and LTR models. We utilised the offline labels as the input for the models and compared the performance of the models with the offline labels. The LTR models did not demonstrate advantages over individual offline labels. On the other hand, GPT-3.5 surpassed both the LTR models and offline labels in successfully placing the MECP in the top position for a given query, showcasing its superior performance in this task when the offline labels were used the the model input. However, we observed that in the absence of the offline labels as the input for GPT-3.5, its performance dropped dramatically. This highlights that despite the recent advancements in LLMs, they are still unable to completely substitute human evaluations in all circumstances.

The impact of query length on the relationship between online and offline evaluations in search clarification is minimal. The evaluation metrics obtained from offline evaluations remain in the same order regardless of query length. However, the highest-performing offline label differs between short and long queries, indicating that different evaluation criteria may be more relevant depending on query length.

**RQ2:** How does uncertainty in the online evaluation impact the relationship between online and offline evaluation?

The reliability of online evaluation data influences the strength of the relationship between online and offline evaluation. When online data is more reliable, a stronger correspondence with offline evaluation is expected. This suggests that ensuring the quality of online evaluation data is crucial for obtaining meaningful insights.

Furthermore, we employed six distinct evaluation metrics and found that the specific choice of metrics can influence the relationship between online and offline evaluations in search clarification. Suppose the goal is to examine both online and offline evaluations to identify the most engaging clarification for a given query. In that case, we suggest focusing on the Precision at Rank 1 (P@1) and Mean Reciprocal Rank (MRR) metrics as top priorities. Metrics such as RBO and RBP that consider binary relevance are inappropriate for comparing online and offline evaluations in search clarification.

Despite the valuable insights provided by this study, certain limitations should be acknowledged. The limitations include:

- It was shown that offline evaluations may not always align fully with online evaluations in certain instances. Enhancing the information given to annotators can improve the consistency between online and offline assessments.
- The study primarily focused on five specific offline evaluation approaches. While these approaches provided valuable insights, other potential methodologies or variations of existing approaches may exist that were not explored in this study.
- The study's findings were based on specific evaluation metrics. Moreover, the observations were based on the experiments conducted on the *MIMICS-Duo* dataset. *MIMICS-Duo* is the only publicly available search clarification dataset containing online and offline evaluations. Larger and more diverse datasets are required to expand the conclusions. The generalisability of the results to other domains or search clarification scenarios also requires additional investigation.

- User engagement is subjective, and users may have varying preferences. While the study considered multiple aspects of user engagement, individual preferences and subjective interpretations of engagement may not be fully captured.

In our study, while acknowledging the potential influence of dataset size, the statistically significant differences we observed in our analysis provide a solid basis for drawing trustworthy conclusions. We have employed rigorous statistical methods to ensure the reliability of our findings, and the observed effects are unlikely to have occurred by chance alone. Based on the conclusions drawn from this study, here are some potential directions for future work:

- Expand and refine offline labels and evaluation metrics: This study focused on five offline evaluation methodologies, but there is room for exploring additional aspects. Future work could also involve developing and testing new evaluation metrics or adapting existing metrics from related fields. This would help in obtaining a more comprehensive understanding of search clarification models.

- Investigate other factors: While the study addressed the impact of query length on the relationship between online and offline evaluation, other factors are worth exploring. Future research could investigate how query intent, topic, or clarity/difficulty influence the relationship between online and offline evaluations. Understanding these factors would provide deeper insights into the effectiveness of search clarification models.

- Apply the Wizard of Oz approach: Conducting experiments using the Wizard of Oz approach [24], where clarification questions are directly asked from users, can provide valuable insights into what factors contribute to making a clarification engaging. This approach involves simulating the functionality of search clarification models through human operators. By studying user interactions and preferences in this setup, researchers can better understand the key elements that make clarifications effective and engaging.

- Improve annotation guidelines: Providing more information to annotators can enhance the correspondence between online and offline evaluations. Future work should focus on developing improved annotation guidelines that provide clearer instructions and examples to annotators. Well-defined guidelines would help ensure consistent and reliable annotations, leading to more accurate offline evaluations.

- Explore other user engagement metrics: We focused on evaluating the effectiveness of search clarification models based on a click-through measurement, future research could explore additional metrics. For instance, sentiment analysis could assess user satisfaction or frustration levels. Integrating such metrics into the evaluation framework would provide a more comprehensive understanding of the impact of search clarification on user experience.

By focusing on these areas of future work, researchers can further advance the understanding of search clarification systems, leading to improved user experiences and more effective communication in various domains.